\documentclass[longauth]{aa}

\usepackage{graphicx}

\begin{document}

\def\be{\begin{equation}}
\def\ee{\end{equation}}
\def\lesssim{\raisebox{-0.3ex}{\mbox{$\,\, \stackrel{<}{_\sim} \,\,$}}}
\def\gtrsim{\raisebox{-0.3ex}{\mbox{$\stackrel{>}{_\sim} \,$}}}
\def\( {\left( }
\def\) {\right) }

   \title{Mimicking neutron star precession by polar cap current-pattern drifting}

     \author{M. Ruderman
          \inst{1,2}
          \and
          J. Gil\inst{2,3}
          }
   \institute{Department of Physics and Columbia Astrophysics
Laboratory, Columbia University, New York, NY\\
\email{mar@astro.columbia.edu}
         \and
Institute of Astronomy, Cambridge, U.K. \and
             Institute of Astronomy, University of Zielona
G\'ora, Lubuska 2, 65-265, Zielona G\'ora, Poland\\
             \email{jag@astro.ia.uz.zgora.pl}
             }

 \date{}
\abstract
{}
 {We propose a model for rotating current patterns within radiopulsar
 polar cap accelerators which has observational consequences that mimic
 those which have been attributed to neutron star precession.}
{The model is a simple extension of a commonly used one for the
origin of the "drifting subpulses" often observed within the pulse
envelope of radiopulsars. }
{The new model's current pattern rotation period (with respect to
the neutron star) is estimated to be of order a year. Associated
with that rotation are small oscillations in spin-down torque,
pulse arrival time, and radiobeam direction with this same period.
These have estimated magnitudes which support a possible
reinterpretation of precession "observations" which could resolve
the problem of obtaining the required precession parameters with
canonical neutron star models.}
{}
\titlerunning{Mimicking neutron star precession}
\authorrunning{Ruderman \& Gil}
 \maketitle


\section{Introduction}

For a considerable number of pulsars the residuals $(\Delta t)$
between observed pulse arrival times and a smooth expectation
based upon average period $P$, $\dot{P}$, and $\ddot{P}$ can be
fit to a single oscillation
\be \Delta t = a\sin\left(\frac{2\pi t}{P_p}\right)
\label{dt}\ee and its overtones (cf Table 1). The physical reality
and significance of this residual oscillation is particularly
compelling for PSR 1828$-$11 (Stairs, Lyne, \& Shemar
\cite{sls00}) whose observed time variation in pulse shape is very
strongly correlated with an oscillating component in $\dot{P}(t)$,
with the same period $P_p$. This correlated pulse-shape component
has discouraged an explanation of the $\Delta t$ oscillation as
neutron star (NS) movement in response to that of planetary
companions. Instead it has generally been considered as strong
evidence that this neutron star has a sustained precession with
period $P_p$ of about 3 years.

\begin{table}
\caption[]{Period $(P_p)$ and amplitudes ($a$) for the oscillating
component in the time residuals between observations and
expectations for several radiopulsars.} \label{t1}
\begin{center}
\begin{tabular}{cccc}
\hline\hline
Pulsar    & $P_p$ (days) & $a(10^{-2}~{\rm sec})$ & Ref \\
\hline Crab & $10^3$ & 1 & LPS88\\
\hline B1828$-$11 & $10^3$ & 2 & SLS00 \\
\hline B1642$-$03 & $2\times 10^3$ & 2 & SLU01\\
\hline B1929$+$10 & $1\times 10^3$ & 1 & U03 \\
\hline B1557$-$50 & $2\times 10^3$ & 0.5 & CUO03 \\
\hline RX J0720 & $3\times 10^3$ & $10^2 (?)$ & HTD06\\
\hline
\end{tabular}
\end{center}
\end{table}

A long-lasting precession period of such magnitude is compatible
with pulsar models which describe a neutron star as a rigid body
(Agk\"{u}n, Link and Wasserman 2006; this paper also contains an
excellent well-referenced survey of previous work on NS
precession.) However, the precession parameters inferred from
observations are much too long to be compatible  with the
canonical models of a NS core containing a neutron-superfluid and
a proton-TypeII-superconductor threaded by a strong magnetic field
linking the core, crust, and external magnetosphere (Link 2003;
see also Shaham 1977, Sedrakian, Wasserman and Cordes 1999, and a
different view by Alpar 2005). In a proton-TypeII-superconductor,
magnetic field is organized into quantized flux-tubes  which
interact strongly with coexisting quantized vortex-lines of the
spinning neutron superfluid. If, in a less standard model, the
protons were to form a Type I-superconductor, a "mixed state"
would appear in which the magnetic field becomes  strong enough in
some regions to quench superconductivity there but vanishes
elsewhere. According to Sedrakian (2005) the drag on vortex-lines
moving inside the very different regions of such a "mixed state"
can be small enough to be compatible with long-period NS
precession. (However, special consideration of expected pinning or
pile-up of vortices at very numerous interfaces between magnetized
and unmagnetized regions still seems needed).

The canonical model for a NS, with its strong interaction between
a core's moving superfluid neutron vortex-lines and the magnetic
field out through the NS's conducting crust, agrees well with a
number of different kinds of pulsar observations: dipole magnetic
field evolution in spinning-down pulsars, magnetic field
structures of millisecond pulsars, Crab-like "glitches", giant
Vela-like "glitches" (Ruderman 2005, 2006). It seems problematic
whether a non-canonical model with a very much weaker interaction
between core vortex-lines and surface magnetic should have similar
consequences.

We propose below models which avoid real and potential conflicts
between  canonical models of neutron star interiors and the Table
1 "precession parameters". In them current-patterns in the pulsar
polar caps (and magnetosphere) rotate in a polar cap $B$-field
which remains fixed in a non-precessing neutron star. The
estimated current-pattern rotation period $(P_p)$ and the
amplitude $(a)$ of the pulse time of arrival oscillation caused by
the current rotation's associated oscillation in NS spin-down
torque are both similar to those in Table~\ref{t1}.

In Section 2 we first review an old model (Ruderman \& Sutherland
\cite{rs75}) for the common rapidly ``drifting subpulses''
observed in many radiopulsars (Backer \cite{b73}; Weltevrede,
Edwards, \& Stappers \cite{wes05}). In it these are a consequence
of $\mathbf{E}\times\mathbf{B}c/B^2$ drifting of patterns of
``sparks'' in polar cap accelerators. The estimated and observed
circulation times of the rotation pattern of these sparks in the
polar cap is tens of seconds. We then consider, in Section~3,
models for a family of related ``drifting spark'' models which
give the very different magnitudes of Table~\ref{t1}.

\begin{figure}
\centering \includegraphics[width=8cm]{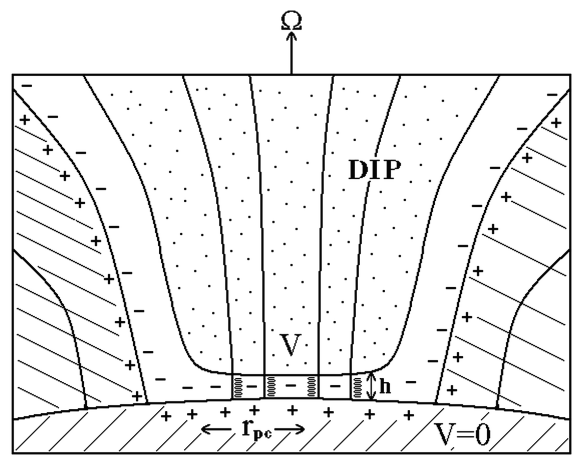}
    \caption{Sketch of an idealized polar cap accelerator.
    The magnetosphere plasma on all closed magnetic field lines corotates with
    the neutron star (hatched regions) separated by an empty ``gap'' from
    a detached inner magnetosphere plasma (DIP). The positive charge $(+)$ which would have existed
    in the gap region before the accelerator formed (Goldreich \& Julian charge density $\rho_{GJ}$)
    is now kept from reentering the gap by the stellar surface (e.g. by ion
    binding to a condensed matter surface) and by  the surrounding open-closed magnetic field
    line boundary where $\mathbf{E}\cdot\mathbf{B}=0$. (Ion inertia alone would have a qualitatively similar
    consequence with ion pile-up on the gap side of the surface boundary). These form an outer surface whose potential is unchanged
    by the insertion of an accelerator $(V=0)$. The inner boundary of the accelerator has a
    constant $V$ ($\sim 10^{12}$ volts) where it is just able to sustain the $e_\pm$ production needed to limit accelerator growth.
    Effectively, the accelerator contains an additional negative charge density $\rho=-\rho_{GJ}$. The inner boundary shape is just that required to make
$\mathbf{E}\cdot\mathbf{B}=0$ at the inner boundary of the
accelerator so there is no charged layer there and no change of
$\textbf{E}$ within the DIP because of the accelerator. In this
model the DIP corotates with the neutron star. The minus signs (-)
indicate the absence of the positive Goldreich \& Julian charge
density need to keep V = 0 everywhere.}
\label{fig1}
\end{figure}

\section{A rotating ``spark'' model for canonical drifting
subpulses}

Deshpande \& Rankin \cite{dr99}, \cite{dr01}) have made the most
detailed observations and analysis of narrow subpulses moving
through pulse envelopes in radiopulsars. Particularly striking is
their deduction of the arrangement and motion relative to the NS
of very many narrow radioemission beams in a single pulsar. In
their analysis of PSR 0943$+$10 they deduced $20$ separate sources
fixed on a moving ``carousel wheel'' rotating in the polar cap's
magnetic field with a period $\sim 40$~sec. Such a moving pattern
has been proposed to come from a similar structure in the current
flow along the open field line bundle of  magnetic field lines
connecting a pulsar's polar cap to its light cylinder. Electric
fields are needed to maintain such outflow. Without such fields,
in the corotating frame the inner magnetosphere's electric field
$(\mathbf{E})$ and velocity $(\mathbf{v})$ could be everywhere
zero. In the laboratory frame
$\mathbf{v}=\mathbf{\Omega}\times\bf{r}$, ${\bf E}={\mathbf
v}\times{\mathbf B}/c$, and the inner magnetosphere would have the
Goldreich-Julian charge density
$\rho_{GJ}=\mathbf{\nabla}\cdot\mathbf{E}/4\pi=\mathbf{\Omega}\cdot\mathbf{B}/2\pi
c$ at all points. Because of current flow through the light
cylinder (and especially when $|\mathbf{\Omega}\cdot\mathbf{\hat
{B}}|$ increases with altitude above the polar cap), more charge
must continually be pulled from the polar cap to maintain that
flow than can be supplied easily. The polar cap surface may not be
an adequate source for the needed particles because of ion
inertia, or perhaps, in some cases, because surface ions are bound
at a condensed matter interface. This extreme case is sketched in
Fig.~1. (If ions can be pulled out but their initial velocities
are suppressed because of ion inertia the positive excess (space
charge) would be inside the accelerator, near its lower
outerboundary. The model's consequences would not be qualitatively
changed. If charge from the NS surface is unavailable, the
$\rho=0$ gap will grow until the additional potential across it
extends to a height $h$ and $e_\pm$ pair production can be
sustained. These pairs are produced by very energetic curvature
$\gamma$-rays from accelerated e$^+$ and e$^-$ in the accelerator.
Copious pair production continues far above the accelerator from
those e$^+$(e$^-$) which flow out from it into that region. Then
because of the consequent high conductivity there,
$\mathbf{E}\cdot\mathbf{B}=0$ in the plasma above the accelerator
where nearly this same $V$ is maintained from the top of the
accelerator to near the light cylinder at $r\sim r_{lc}\equiv
c/\Omega$. The indicated potentials and charges in Fig.~1 refer
only to the differences from those where $\rho=\rho_{GJ}$
everywhere. The NS and all plasma on closed field lines (where
$\rho$ still equals $\rho_{GJ}$) still corotate. They are bounded
by the $V=0$ surface. The upper surface of the empty accelerator
zone has a fixed potential $V$ and a position and shape such that
$\mathbf{E}\cdot\mathbf{B}=0$ as well as $\mathbf{E}$ tangential
vanishing on it in the corotating frame. Then, except at the
corners where the accelerator's outer boundary bends suddenly
upwards along the last open field line, the accelerator thickness
\be h\sim \left(\frac{V}{2\pi\rho_{GJ}}\right)^{1/2}. \label{h}\ee

The entire detached inner magnetosphere plasma (DIP) above the
accelerator is copiously filled with e$^\pm$ pairs and it is
assumed to have this same $V$ throughout it. It does not have any
change in its initial corotation ($\mathbf{E}\times\mathbf{B}=0$)
with the NS. In this model the corotation is broken only within
the accelerator where a charged particle will move with respect to
the NS with an angular speed about a polar cap axis \be
\Omega_p=\left|\frac{\mathbf{E}\times\mathbf{B}}{B^2}\right|\frac{c}{r_\perp}\sim\frac{Vc}{hB
r_\perp}|\mathbf{\hat{E}}\times\mathbf{\hat{B}}|. \label{omega}\ee
$|\mathbf{\hat{E}}\times\mathbf{\hat{B}}|$, the sin of the angle
between accelerator $\mathbf{E}$ and NS $\mathbf{B}$ is $0$ at the
model polar caps axis $(r_\perp=0)$ and $\sim 1$ at the corner
where $r_\perp\sim r_{pc}$ (radius of polar cap) and in the slot
above it (see Fig.~1 for illustration).

With the very crude extrapolation $|\hat{E}\times\hat{B}|\sim
r_\perp/r_{pc}$, equation~\ref{omega} becomes \be
\Omega_p\sim\frac{Vc^2r_{pc}}{hB_{d} \Omega R^3}.
\label{omegap}\ee This estimate for $\Omega_p$ is not sensitive to
spark location (``slot'', near the periphery $r_\perp\sim r_{pc}$
or much closer to the $r_\perp=0$ axis), and we indicate locations
of 4 exemplary sparks in Fig.~1. If the current flow pattern
through the accelerator consists of particle flow in both
directions bunched into ``sparks'' separated by a distance $\sim
h$, a stable pattern of such sparks (see Beskin 1982, Gil \&
Sendyk 2000) could rotate at the equation~(\ref{omegap}) rate
(e.g. returning spark particles heat a surface spot which
determines the next preferred emission spot, etc.).

For the pulsar parameters of B1828$-$11, $P=0.4$~s, dipole
$B_d=5\times 10^{12}$~G, ``characteristic spin-down age'' $10^5$
yrs, polar cap radius $2\times 10^4$~cm and accelerator thickness
$h\sim 6\times 10^3$~cm (chosen quite arbitrarily to give some
partial filling of the otherwise empty accelerator gap because of
ion flow into it; e.g. Gil, Melikidze \& Geppert 2003) and $V\sim
10^{12}$ volts, $P_p=2\pi/\Omega_p\sim 40$~sec and the number of
circulating sparks on each circulating carousel wheel of the
sparks pattern
\be N_s\sim\frac{2\pi r_{pc}}{h}\sim 20 .\label{Ns}\ee

The good agreement with $\Omega_p$ and $N_s$ deduced by Deshpande
\& Rankin \cite{dr99} for PSR B0943+10 is comforting but far from
compelling and very different kinds of models have also been
proposed for such drifting subpulse observations in which
outer-magnetosphere plasma plays an important role (cf. Wright
2003). Weltevrede and colleagues (2005) have shown the lack of
correlation between observed circulation periods of drifting
subpulses and various other pulsar observables. Moreover, there is
often great uncertainty about inferring needed polar cap
accelerator parameters from those observables. [For example, in
several pulsars, and probably in a large fraction of them, the
polar cap magnetic field seems to be very much larger (and the
polar cap area very much smaller) than the surface dipole field
inferred from the pulsar's spin-down torque]. A relatively robust
prediction of the drifting polar cap model of equations (4) and
(5) comes from combining them to give a predicted angular rotation
speed of the polar cap's spark-determined current pattern
independent of the choice of polar cap radius or accelerator
height, but depending only on the number of sparks in the
carousel. For PSR 0943+10, where that number has been determined,
the agreement between model and observed carousel spin-rate is
good. In summary, models  based upon carousels of polar cap
sparks, predict carousel rotation periods (of order tens of
seconds) and numbers of sparks which are not inconsistent with
drifting subpulse observations.

\section{Very long period circulations of polar cap currents and their consequences}

The carousels  discussed above rotate because accelerated
particles from a polar cap surface make pairs, one member of which
returns to the surface at a point displaced from the release point
of its parent. The next generation repeats this process, but
begins from the new origin. Carousel rotation periods of tens of
seconds occur because the time it takes an electron (positron)
from a pair created within the accelerator to return to the polar
cap surface is the same as the time it spends "drifting" in the
strong electric field of the accelerator. Rotation periods up to
$10^6$ times longer than those considered above would result if
the pairs were first separated far above the polar cap region (
for example, in the outer-magnetosphere near the light cylinder).
The time of flight back to the polar cap surface for a backflowing
member of that pair would typically be huge relative the time
during which it feels the strong drift fields in the relatively
small polar cap accelerator.

The processes which make pairs within a polar cap accelerator
continue to make them well above the accelerator. There is a large
pair flux flowing up the open field line bundle from the DIP out
to the light cylinder. Associated with it would be a small
potential drop, $V'$ of several $mc^2/e\sim 10^6$ volts, since an
electric field is needed continually to adjust local net $\rho$ to
$\rho_{GJ}$ because of growing
$\left|\mathbf{\Omega}\cdot\mathbf{\hat{B}}\right|$ along the pair
flow. One way of adjusting the net charge density (generally a
very small fraction of the total pair density) is by the reversal
in flow direction of relatively few electrons (positrons) which
then flow back toward the NS surface. The needed readjustment in
charge density could well be accomplished in other ways. In the
absence of enough quantitative knowledge of the open field line
electric field along B out to very near the light-cylinder we
simply assume some backflow. If this backflow dominates that
directly onto the polar cap surface from the e$^\pm$ pairs created
and separated within the accelerator, the RHS of
equations~(\ref{omegap}) and (\ref{Ns}) should be decreased by
about $h\Omega/c$. Then \be \Omega_p\sim\frac{\Omega V}{B
r_{pc}}\sim\frac{V}{B R}\left(\frac{c \Omega}{R}\right)^{1/2}
.\label{omegap2}\ee To it should be added an additional rotation
from $V'$ with \be \Omega_p^\prime\sim\frac{V'c^2}{R^3B\Omega} .
\label{omegap3}\ee With PSR B1828$-$11 parameters,
$\Omega_p^\prime\sim 10^{-1}\Omega_p$ for peripheral and interior
spark patterns. For slot gap sparks the gap accelerators keep
about the same $\Omega_p$ all the way up the slot so that
equations~(\ref{omegap}) and (\ref{Ns}) remain valid for them. For
PSR B1828$-$11 equation~(\ref{omegap2}) gives $P_p\sim 0.5$~yr.
For the Crab pulsar $P_p\sim 0.1$~yr.

The current distribution pattern on the polar cap and the ${\bf
j}\times {\bf B}$ forces will determine the large Goldreich-Julian
torque on a pulsar. We assume that associated with a polar cap
current-pattern rotation should be a small oscillatory component
with the same period in the pulsar's spin-down torque;
$\delta\dot{\Omega}=\propto\dot{\Omega}\sin(\Omega_pt)$. This
would give an amplitude in equation~(\ref{dt}) of \be
a=\frac{\alpha}{2t_{sd}}\left(\frac{P_p}{2\pi}\right)^2
\label{a},\ee with $t_{sd}$ the characteristic spin-down age
$(P/2\dot{P})$. We note how much larger $a$ becomes as $P_p$
increases from the 40 s of equation~(\ref{omegap}) to $\sim 10^7$
sec in equation~(\ref{omegap2}).

A rough estimate of the dimensionless $\alpha$ is the product of
the relative area of any expected rotating irregularity in the
rotating current pattern $(h^2/r_{pc}^2)$ times the fractional
effect on the Goldreich-Julian spin-down torque from moving that
current from one side of the polar cap to the opposite
$(r_{pc}/R)$, times the fraction $f(\sim 1/2)$) of the spin-down
torque which comes from Goldreich-Julian polar cap currents. Then
$\alpha\sim h^2/2r_pR$ and, for PSR B1828$-$11 $a\sim 3\times
10^{-2}$~sec. For the Crab pulsar with the same $V$, $a\sim
0.4\times 10^{-2}$~sec. The $\Delta t$ of equation~(\ref{dt})
$\propto B^3\Omega^{-1/2}$ so that predicted changes in $\Delta t$
are very much less than the differences of $10^2$ between the Crab
and PSR B1828's $t_{sd}$\footnote{With realistic polar cap
$\mathbf{B}$ $\gamma$-rays produced near a NS by a polar cap
accelerator will be gravitationally bent to become a strong source
of e$^\pm$ creation in large parts of the nearby magnetosphere.
Magnetic field structure can have a similar consequence. These
would give e$^\pm$ flows on open field lines which are not
necessarily separated from the polar cap by the accelerator
potential $V$. Indeed, such pair flow might even quench parts of
that accelerator. However the argument for a $V'$ would still
hold. $\Omega'_p$ would still be present even in the absence of
any $\Omega_p$ for the involved $e^-/e^+$. Then for PSR B1828$-$11
$P'_p=4$ yrs and for the Crab $P'_p=50$ yrs.}. An exceptionally
large amplitude would be expected and is observed for RXJ0720.

\section{Summary}

Our proposed model is a very simple extension of one commonly used
to describe the origin of canonical ``drifting subpulses'':
${\mathbf E}\times{\mathbf B}$ drift of localized $e_\pm$ sparks
within a polar-cap accelerator. It differs only in the time
$e_-(e^+)$ coming back to the stellar surface has spent inside the
accelerator ($h/c$, about $10^{-7} s$) relative to the time spent
above it ($0-\Omega^{-1}$, about $0-10^{-1}$~s).  If the latter
time $= 0$ we have the old drifting subpulse model which gives
carousel circulation times of some 10's of seconds; if it is near
the maximum time of flight back from near the light cylinder, we
have the model of section~3. (The choice between them may depend
on the detailed structure of the polar cap's surface magnetic
field.) The very rough estimates of section~4 for $P_p$ and
$\Delta t$ are similar to those of Table 1. While all of this is
hardly a compelling argument for the kind of model proposed here
it does seem to support  further consideration as a not
implausible way of resolving the problem of how a canonical
neutron star could sustain the small precession, which has been
widely suggested as the cause of very long period oscillations in
timing observables of some pulsars.
\begin{acknowledgements}
We acknowledge the support of the Polish State Committee for
scientific research under Grant 1 P03D 029 26. Both authors are
especially grateful to the Institute of Astronomy in Cambridge for
its hospitality and to the anonymous referee for very helpful
suggestions.

\end{acknowledgements}

{}

\end{document}